# 3-PHASE RECOGNITION APPROACH TO PSEUDO 3D BUILDING GENERATION FROM 2D FLOOR PLAN

Raj Kishen Moloo

Computer Science and Engineering Department,
University of Mauritius, Reduit, Mauritius
r.moloo@uom.ac.mu

Muhammad Ajmal Sheik Dawood

MSc Student, University of Essex, Colchester, Essex, UK
ajmalsd@yahoo.com

Abu Salmaan Auleear

Senior Web-Designer, Chesteroc ltd (Mundocom Mauritius), Ebène, Mauritius
prince_05@live.com

## ABSTRACT

*Nowadays three dimension (3D) architectural visualisation has become a powerful tool in the conceptualisation, design and presentation of architectural products in the construction industry, providing realistic interaction and walkthrough on engineering products. Traditional ways of implementing 3D models involves the use of specialised 3D authoring tools along with skilled 3D designers with blueprints of the model and this is a slow and laborious process. The aim of this paper is to automate this process by simply analyzing the blueprint document and generating the 3D scene automatically. For this purpose we have devised a 3-Phase recognition approach to pseudo 3D building generation from 2D floor plan and developed a software accordingly.*

*Our 3-phased 3D building system has been implemented using C, C++ and OpenCV library [24] for the Image Processing module;  The Save Module generated an XML file for storing the processed floor plan objects attributes; while the  Irrlicht [14] game engine was used to implement the Interactive 3D module. Though still at its infancy, our proposed system gave commendable results. We tested our system on 6 floor plans with complexities ranging from low to high and the results seems to be very promising with an average processing time of around 3s and a 3D generation in 4s. In addition the system provides an interactive walk-though and allows users to modify components.*

## 1. INTRODUCTION

Enhanced graphic processing capabilities of today's computers have enabled the widespread of three dimension (3D) visualisation in the construction industry, providing realistic interaction and walkthrough on engineering products. 3D architectural visualization has become a powerful tool in the conceptualisation, design and presentation of architectural products [8]. Long gone are the days of mere 2D plans. Several 3D authoring packages like Google Sketchup [10], 3DS Max [11], Maya [12], Blender [13] can be used to model 3D buildings based on the 2D plan. Nevertheless, creating 3D buildings requires skilled and trained people, besides being a lengthy, cumbersome and tedious task [8].

Nowadays, popular architectural package like AutoCad [9] allows saving of 2D plans digitally, in particular formats, e.g., DWG/ DWF format, out of which enhanced design and visualization workflows on 3D buildings can be generated. However, not every building plan has been implemented using these advanced engineering software. There are still plans which are hand-drawn or are not saved in particular structured format allowing 3D generation. The aim of our paper is to tackle this issue whereby no digital 2D plan exists (e.g. DWG format), accepting a 2D floor plan, scan it as an image, pre-process the plan and reconstruct a 3D plan out of the 2D plan. For this purpose, image processing techniques have been applied to extract semantic and spatial information, recognising architectural symbols such as walls, windows and doors only and locate their positions in the 2D sketch so that these details are used to generate a pseudo-3D model of the building. Information extracted is stored in an XML format which can be easily ported to any architectural package format. Based on our XML format we used a game engine, Irrlicht [14], to generate the 3D plan allowing navigation and walkthrough.

Several work on this theme have been done previously as elaborated in section 2 (Previous work), with research spanning from effective symbol recognition algorithm to implementation of whole system/package to generate 3D models out of 2D floor plans. The aim of our paper is to implement the same step, but using existing image processing library like OpenCV [24] and 3D engines. Functions provided by image processing library OpenCV 2.2 [24] have been used throughout this project to evaluate and find the most appropriate image processing techniques that can be used for the detection and extraction of the architectural symbols. Only walls, windows and doors symbols which are oriented horizontally and vertically have been considered while building the system. Also, once pre-processed, spatial and semantic information are gathered and stored in XML format to be loaded in an existing 3D engine. Though our system is at its infancy and does not cater for all possibilities, interesting results have been noticed in terms of symbol detection rate and 3D building generation.

The rest of the paper is organised as follows: Section 2 describes similar existing work on architectural symbol recognition and 3D building generation. Section 3 provides a description of our proposed system while in Section 4, we elaborate on our implementation process. Section 5 analyses the results achieved through our system. Finally, we conclude in terms of our achievements and future directions in Section 6.

## 2. PREVIOUS WORK

Lewis and Sequin in [1] developed a robust, semi-automatic system to create 3D polyhedral building models from computer drawn floor plans, requiring minimal user interaction. A prototype system, called the Building Model Generator (BMG) which accepts 2D floor plans in the common AutoCAD DXF geometry description format. The system starts by cleaning up the 2D floor plans, then converting them into suitable internal data structure permitting efficient geometric manipulations and analysis. After correction of small local geometrical inconsistencies and necessary adjustments are made, a consistent layout topology is obtained. Semantic information such as room identities and connecting portals are extracted, walls are extruded to a specified height, and door and window geometries are inserted where appropriate. Hence a 3D representation of the building is generated permitting visualisation in interactive walkthrough. The models generated are directly compatible with the Berkeley WALKTHROUGH system and with the NIST CFAST fire simulator. Such work demonstrated that the construction of 3D building models from existing floor plans have become manageable and affordable reducing built-in time from months to days.

Lu & al in [2] proposed new methods to analyse architectural working drawings (AWD), usually used to describe design intents of architects, recognize typical structural objects and architectural symbols. They categorise AWD into three types of entities namely structural, functional and decorative. Structural entities deal with structures supporting and resisting load of the whole building such as steel and concrete. Functional entities such as doors, windows, lifts etc provide convenience to the users. Examples of decorative entities are partition walls, hung ceilings etc. Processing of these AWDs are done in three steps namely recognition of structural entities using shape based method, removal of graphical primitives of recognized structural objects and recognition of architectural entities from simplified drawings using feature-based symbol recognition. They test their model on 257 plane architectural/structural drawings, 40 being synthetic and 217 being real data. This model shows interesting detection rate with above 80% correct detection rate on synthetic and real data set.

S.-H. Or et al in [3] proposes a highly automated approach to generate 3D model from a 2D floor plan. The system parses a floor plan into a number of connected segments and analyse their relationship to generate the 3D model accordingly. The plan is pre-processed converting the raster image into vector image. Symbol recognition is used for identification of doors and windows. The system assumes that any arc symbols represent a door while thin boxes represent windows. Once these symbols are identified, the 3D models are generated and loaded on Genesis 3D, a 3D game engine. However, the system needs some improvements in terms of recognition of building entities, creation of multi-storey building.

Dosch et al in [4] proposed a complete system to reconstruct scanned 2D architectural drawings in 3D. They describe their robust image processing and feature extraction algorithm by dividing the scanned 2D images into tiles, processing each part separately and then merging them again after vectorisation. Using the skeleton based approach they extract lines and represent them into segments using the polygon approximation technique. Their system allows for dashed line, arcs and staircase detection. A user interface is provided for human interaction and correction of errors and interactively manipulating the resulting data. They advocate the stability of their image processing and feature and dashed line extraction algorithm. Nevertheless, there are still some improvements need to be done and tests need to be made on how scalable this system is.

[5] proposed a Self-Incremental Axis-Net-based Hierarchical Recognition (SINEHIR) model for automatic recognition and interpretation of real-life complex electronic construction structural drawings. They designed and implemented shape-independent algorithms based on internal semantic constraints rather than visual graphical constraints. Their methods identifies characteristic features of structural components from the more regular constituents, and then tracks the graphic objects as far as possible under the guidance and constraints of recognised objects and the domain knowledge. Their approach was tested on more than 200 real-life drawings and the results showed a 90% average recognition rate.

In [6], based on Messmer & Bunke ideas on symbol recognition, Ah-Soon & Tombre proposed a method for recognising architectural symbols based on the description of the model through a set of constraints on geometrical features, and on propagating the features extracted from a drawing through the network of constraints. They advocate the flexibility of such approach to accommodate new symbols.

[7] proposes a reconstruction method based on four main phases namely (i) 2D edges processing for removing geometrical inconsistencies, (ii) topological reconstruction with semantic information, (iii) 3D building extrusion (iv) superimposing of floors. The model expresses incidence and adjacency relations between all the elements with semantic information associated with all volumes such as openings, portals, stairs etc. Conversion from 2D to 3D is done through extrusion using specific rules guided by the semantics.

# 3. PROPOSED SYSTEM

Our proposed system is a 3-phase system namely (i) Image Processing (ii) Save Module (iii) Interactive 3D module.

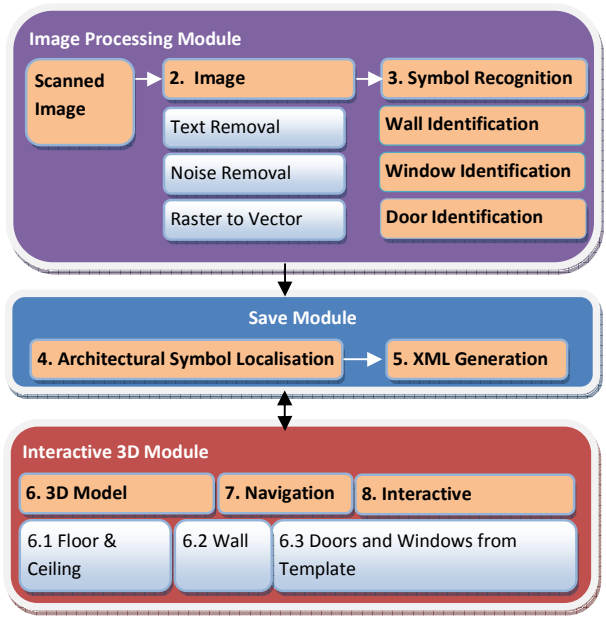

Figure 1. 3-Phase proposed system

## 3.1 Image Processing

The system loads a scanned image of a floor plan consisting of wall, door and window symbols oriented horizontally and vertically. The scanned image is pre-processed by removing noise and text, converting it to a vectorised image. The image is preprocessed through thresholding and edge detection. Horizontal and vertical lines are extracted from the resulting image using probabilistic Hough transform. A region growing approach is then applied to cluster the lines to identify walls and derive their positions. Then, the windows are identified from the detected walls using the characteristics of the wall symbols and using a similar region growing approach to walls' identification. The detected walls and windows are subtracted from the image to obtain a difference image with only door symbols. The difference image is then segmented using a contour finding approach to locate the position of each object in the image. A histogram matching technique is applied to identify which of the remaining objects is a door.

### 3.2 Save Module

For each recognised door, their relative position in the image is calculated. Finally the dimension of a pseudo roof for the building is calculated from the sketch. All of the identified walls, windows, doors and approximate roof positions are saved in an XML file.

### 3.3 Interactive 3D module

The Interactive 3D generation module calculates the position, length, height and width of the wall, door and windows models from the generated XML file. The 3D model of the building is then generated, enabling the user to visualise the building in different perspective and navigate through it. The system provides a user friendly menu driven interface. Additional facilities that the system provides are (i) full screen navigation, (ii) interactive modification of 3D models, e.g., applying different textures to either interior or exterior walls separately, (iii) loading different window, door models, roof and tiles, (iv) taking screenshots while navigating, (v) saving XML containing the modified 3D building description and (vi) loading of existing XML file.

## 4. IMPLEMENTATION

The object detection and localisation module of the system has been implemented in C/C++ using data structures and functions provided by OpenCV image processing library. TinyXml was used as an XML parser with Irrlicht Engine to create the GUI, 3D generation and navigation system.

### 4.1 Image Preprocessing

### 4.1.1 Wall Identification

To identify walls, lines are extracted from the preprocessed image using probabilistic Hough transform [20] resulting in a sequence of lines. The lines are then separated as horizontal and vertical lines sequences only. For each sequence of lines, a clustering technique is applied to group lines into bounding boxes, each representing a potential wall. The bounding box is dynamically adjusted to accommodate inserted elements. The number of horizontal bounding boxes is reduced by any grouping overlapping boxes as one single bounding box. Similar technique is used to group vertical lines in bounding boxes to identify potential vertical walls. Intersecting endings of vertical and horizontal bounding boxes are then aligned as shown in the *Figure 2* below. It should be noted that the identified walls may contain windows. The clustering algorithm assumes that for a line forming part of wall in the image and to be included in a bounding box, it should be at most 50 pixels from the edge of the bounding box. This value is used as the stopping condition for the clustering algorithm. Considering horizontal sequence of lines only, the following algorithm identifies potential horizontal walls:

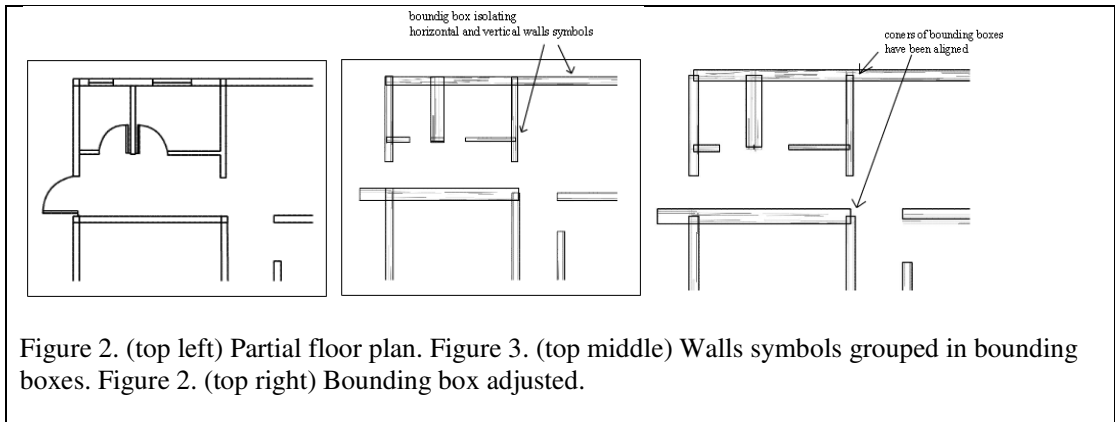

Figure 2. (top left) Partial floor plan. Figure 3. (top middle) Walls symbols grouped in bounding boxes. Figure 2. (top right) Bounding box adjusted.

```
WHILE (horizontal sequence, h, contains lines)
        Put first line from h in bounding box
        FOR each line in h
                IF (line is in (bounding box enlarged by 50 pixels on each side)) THEN
                        Put line in bounding box
                END IF
        END FOR
        FOR each line in bounding box
                Remove line from h
        END FOR
END WHILE
```

Figure 3. Algorithm for bounding box.

## 4.1.2 Window Identification

The same procedure as wall identification, before alignment of horizontal and vertical boxes, has been used for window detection. The algorithm in *Figure 4* describes the extractions of vertical window symbols. Here we assumed that the length of *line_v* should be at least three-quarter (3/4) height of bounding box to be accepted. Height of bounding box must be at two (2) times longer than width of Region of Interest (ROI) and less than half height of ROI to be considered a window. Windows cannot be found at corners of ROI i.e., wall symbols.

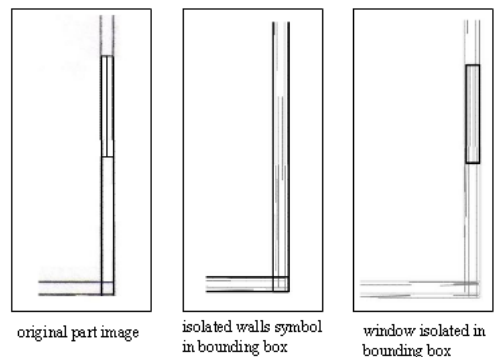

original part image | isolated walls symbol in bounding box | window isolated in bounding box

Figure 4. Wall Identification

```
FOR each vertical bounding box positions
        Set positions as ROI in preprocessed image
        Get width and height of ROI
        Extract lines L, from in ROI using probabilistic Hough transform function
        Get horizontal lines h, from L using getHorLines method
        Get vertical lines v, from L using getVerLines method
        Sort h using sortTopDown method
        Sort v using sortLeftRight method
        FOR each line in h, line_h
                Put line_h and next line_h in bounding box
                FOR each line in v, line_v
                IF (length line_v ≥ 0.75 × height of bounding box) THEN
                        Put line_v in bounding box
                END IF
        END FOR
        IF (height of bounding box ≥ (2 ×width) AND ≤ height÷2) THEN
                Increment number of windows (bounding boxes)
        ELSE
                Clear bounding box
        END IF
END FOR
```

Figure 5. Window symbol detection

### 4.1.3 Door Identification

After walls and windows were identified from a floor plan image, these symbols were subtracted from the edge image leaving only the remaining objects which were then segmented and put in bounding boxes. *Figure 6* shows the isolation of remaining objects before door identification.

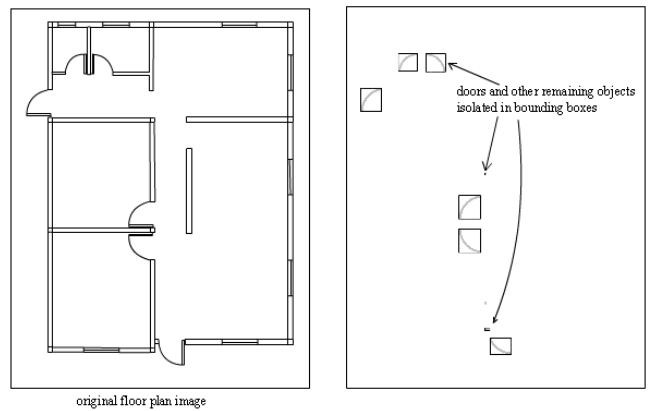

Figure 6. Isolation of remaining objects

Each identified object then undergoes histogram matching which used a combination of Chi-square and Bhattacharyya distance metrics, and template matching which uses normalised correlation coefficient matching method, against a small set of sample door images. After this operation, only bounding boxes containing door symbols are left. It can be said that door symbols detection has a very high success rate.

During histogram matching, an assumption was made that for a ROI to be detected as a door symbol, the results by both Chi-square and Bhattacharyya distance metrics must be below than 0.2, i.e., a good match. The assumption that was made for template matching template is that the result obtained by using correlation coefficient method should be above 0.9 to be a good match. If either histogram or template matching is successful, then the ROI is classed as a door symbol. The implementation details to detect doors is describes in *Figure 7*.

```
Number of samples for templates, TS.
Number of samples for histogram matching, HS.
FOR each vertical wall positions
        Subtract vertical wall image part from image_edge
END FOR
FOR each horizontal wall positions
        Subtract horizontal wall image part from image_edge
END FOR
Find contours in image_edge using cvfindContours function
Redraw on image_edge using redraw method
Find contours in image_edge
FOR each set of contours
        Put contour in bounding box, b
        Increment number of bounding box
END FOR
Set isDoorHistogram to false
Set isDoorTemplate to false
FOR each ROI represented by array b
        FOR i:=0 to HS
                Load sample door image i
Do histogram matching of ROI and sample using compareImgHistogram method
                IF (match is true) THEN
                        Set isDoorHistogram to true
                END IF
        END FOR
        FOR i:=0 to TS
                Load template door image i
        Do template matching of ROI and template using compareImgTemplate method
                IF (match is true) THEN
                        Set isDoorTemplate to true
                END IF
        END FOR
        IF ((isDoorHistogram OR is DoorTemplate) is true) THEN
                Get door bounding box positions
                Increment number of doors
        END IF
END F
```

Figure 7. Algorithm for door identification

## 4.2 Getting door and window positions

After the doors have been identified, they need to be aligned to their adjoining walls. This is achieved by incrementing each side of the bounding boxes representing doors by the average thinness of walls in the image and calculating which horizontal or vertical walls, previously detected, ends in the bounding box. Thus the start and end points for the doors, being represented as lines, are obtained.

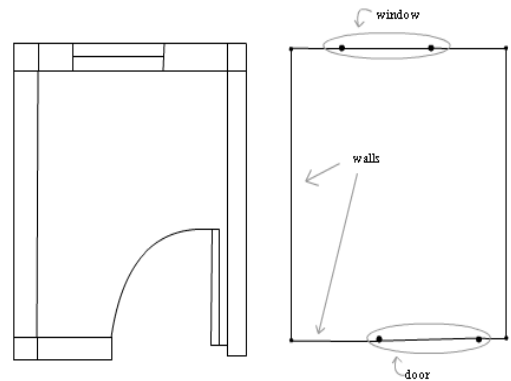

Figure 8. Aligning window to adjoining wall

## 4.3 Saving detected object

After symbols identification, the detected walls, windows, doors [and roof positions] are saved in the XML file. *Figure 9* shows a partial xml contain a floor plan attributes.

```
<?xml version="1.0" ?>
<building>
    <!-- objects and dimensions in  one floor building -->
    <wall w_id="0" ltexture=" " rtexture=" ">
        <point>
            <x1>124</x1>
            <y1>37</y1>
            <x2>1210</x2>
            <y2>37</y2>
        </point>
    </wall>
    <wall w_id="0" ltexture=" " rtexture=" ">
        <point>
            <x1>124</x1>
            <y1>253</y1>
            <x2>200</x2>
            <y2>253</y2>
        </point>
    </wall>
    <wall w_id="0" ltexture=" " rtexture=" ">
        <point>
            <x1>404</x1>
            <y1>253</y1>
            <x2>583</x2>
            <y2>253</y2>
        </point>
```

Figure 9. Partial XML file generated

## 4.4 User Interaction and Generating the 3D model

The graphical user interface (GUI) and the 3D generation module have been implemented in C/C++ using the Irrlicht engine. The GUI allows the user to load a floor plan image and trigger the object detection module which produces an XML file with the detected objects' attributes. The latter is then read by the 3D generation module which calculates the coordinates at which to place the objects such as the walls, doors and windows in the 3D model. It then loads the corresponding objects, map their textures draws the 3D model of the building. The GUI presents the 3D model of the building in several different views as shown in *Figure 10-11* below.

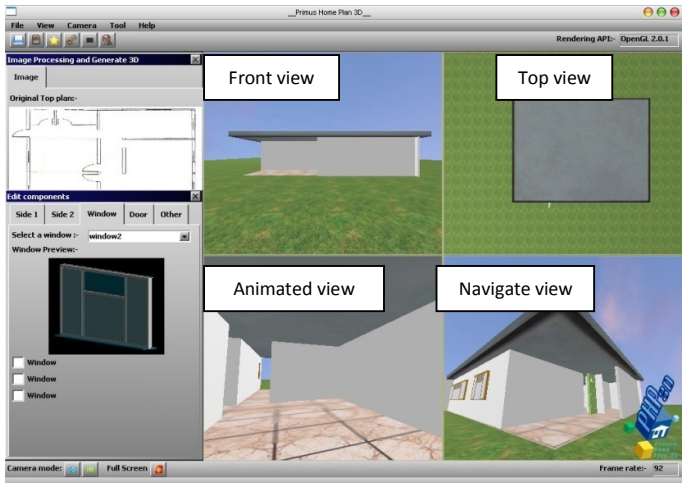

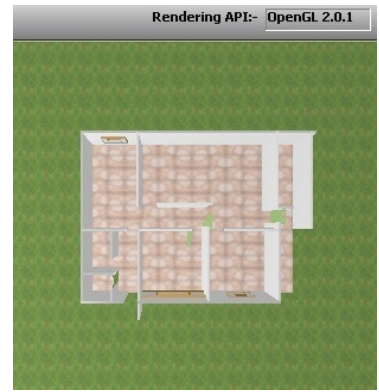

Figure 10. Generated 3D house

Figure 11. Top view without the roof

# 5. RESULTS AND ANALYSIS

Our 3-phased 3D building system has been implemented using C, C++ and OpenCV library for the *Image Processing module*; The *Save Module* generated an XML file for storing the processed floor plan objects attribute; while the Irrlicht [14] game engine was used to implement the *Interactive 3D module*. Though still at its infancy, our proposed system gave commendable results.

## 5.1 Benchmark Testing

In this section, testing is made to know the performance of the software on different hardware systems. The frame rate of the software has been considered for testing the performance.

Table 1. Capabilities of system tested

| System A: | System B: |
|---|---|
| 32MB graphics card | 256MB graphics card |
| 256Mb of RAM | 1GB of RAM |
| 750 MHz Processor | 2.2 MHz Processor |
| Frame rate: ~16 | Frame rate: ~250-300 |

## 5.2 Portability Test

The software was compiled and run using Codeblocks IDE on Fedora 10, a Linux operating system and it could successfully load an image and generate the 3D model, though the frame rate was very low (frame rate was 4 as the proper graphic card driver was not installed on the OS).

## 5.3 Performance Test

Table 2. Performance Testing results

|  | Floor plan 1 | Floor plan 2 | Floor plan 3 | Floor plan 4 | Floor plan 5 | Floor plan 6 |
|---|---|---|---|---|---|---|
| Complexity | Low | Medium | Medium | Low | High | Medium |
| Wall detection | 25 /25 - 100% | 24/30 – 80% | 34/54 – 62.9% | 17/17 - 100% | 38/49 – 77.6% | 26/30-86.7% |
| Window detection | 8/9 - 88.9% | 3/9 – 33.3% | 2/8 - 25% | 3/6 - 50% | 2/12 – 16.7% | 2/4 - 50% |
| Door Detection | 6/7 – 85.7% | 6/14- 42.8% | 4/17 - 23.5% | 8/8 - 100% | 9/15 - 60.0% | 5/9 – 55.6% |
| Image Processing time | 2 s | 3s | 3s | 2s | 5s | 4s |
|  |  |  |  |  |  |  |
| **3D model** |  |  |  |  |  |  |
| Faces | 246 | 355 | 535 | 203 | 511 | 427 |
| Triangle mesh | 884 | 1210 | 1590 | 758 | 1528 | 1409 |
| Average generation time | 4s | 4s | 4s | 4s | 4s | 4s |

Based on *Table 2* it can be said that the system produced commendable results in recognition of walls, windows and doors symbols from floor plans images. Walls positions were extracted with a very high success rate. Windows identification using probabilistic Hough transform and clustering technique produced fairly good results. The combination of histogram and template matching to identify door symbols was very reliable as it allows all doors to be detected in floor plan images. However, better algorithms could not be derived to properly align all the detected doors in their final positions.

## 5.4 Processing Time

Image processing for identification of objects depended on the complexity of the floor plan. The more complex the plan in terms of walls, doors and windows, the more time it took for processing. Computing times have been obtained with a 2.2 MHz processor, 1GB RAM and 256 MB RAM graphic card. However, once the XML was generated, then the 3D model generation of the Building was performed on average in 3 seconds irrespective of the model.

# 6. CONCLUSION

In this paper we demonstrated some conclusive results on our 3-phase recognition approach to pseudo 3D building generation from 2D floor plan. Our system based itself on the idea of using existing concepts, algorithm, libraries and technologies with minimal modifications and integrating them into a fully functional working application. Hence, we used C, C++ and OpenCV for image processing and recognition, XML for storage and Irrlicht [14] engine for 3D model generation.

We used existing image processing techniques like probabilistic Hough transform, Histogram matching, Clustering and bounding box techniques , Chi-square and Bhattacharyya distance metrics to recognise architectural symbols like walls, doors and windows. XML technology was used as the interface and the bonding technology between image processing and 3D generation of building. It was also used as the storage format for saving models and processed image. The 3D Generation module provides a sense of aesthetic and a realistic scene in the 3D environment for the user's experience with the system. It provides interactive and navigational capabilities with different views.

However, the system developed is not 100% efficient in identifying and localising the walls, windows and doors within the floor plans used during the tests carried out. Nevertheless it has been able to detect most wall symbols with some tiny false positive walls forming part of a door symbols being recognised as part of a wall; most window symbols are detected and door symbols have been recognised. However, some doors are not properly aligned with their adjacent walls after being detected and this need some improvement. Also, our 3D module is still at an embryo stage. It needs to be enhanced to produce more realistic model, enabling users to change color, textures, light and the ability to interactively insert objects like tables, chairs etc in the model generated.

## 6.1 Future Works

Though the object detection and localisation module is fully functional some feasible improvements can done to further enhance the system. Some identified future works are as follows: (1) recognition of walls other than at horizontal and vertical positions. This would involve replacing the rectangular bounding box used in the module by a bounding box which can be inclined at different angles; (2) detection of different types of doors and windows such as sliding, interior or exterior doors by modifying the template matching method; (3) detection of other architectural symbols such as supporting columns and stairs; (4) extending the system to allow for multiple storey building.

As a concluding note it can be said that using our approach based on architectural symbol recognition, significant reduction in time is to be noted in 3D building generation as compared to traditional 3D building implementation. This has opened new avenues with commercial potential in creating virtual environments especially for games, virtual property advertising and also for modelling historical building having only paper plans.

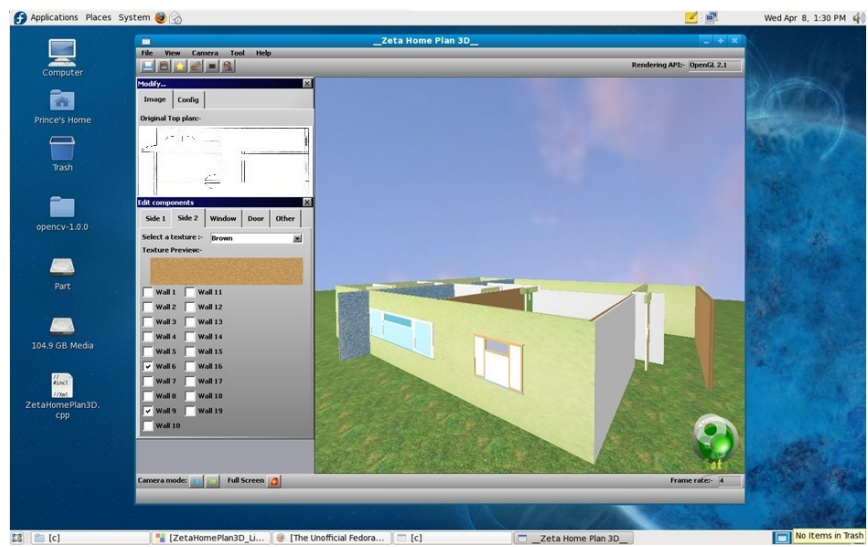

Figure 12. Overview of generated 3D model

# REFERENCES

[1] R.Lewist, & C.Sequin. (1998). Generation of 3D building models from 2D architectural plans. *Computer Aided Design.*30(10), pp. 765-779. Elsevier Science.

[2] Lu, T., Yang, H., Yang, R., & Cai, S. (2006). Automatic analysis and integration of architectural drawings. *Internation Journal on Document Analysis and Recognition* IJDAR, 9 (1)pp. 31-47.

[3] S H Or, et al. (2005). Highly Automatic Approach to Architectural Floorplan Image Understanding and Model Genration. *Proceedings of Vision, Modeling, and Visualization,Pattern Recognition* (pp. 25-32). Erlangen,Germany: IOS Press.

[4] Dosch, P., Tombre, K., Ah-Soon, C., & Masini, G. (2000). A complete system for the analysis of architectural drawings. *International Journal on Document Analysis and Recognition* , 3(2) pp.102-116.

[5] Lu, T., Taib, C.-L., Sua, F., & Cai, S. (2005). A new recognition model for electronic architectural drawings. *Computer Aided Design* , 37(10) pp. 1053-1069.

[6] Ah-Soon, C. et al., (2001). Architectural symbol recognition using a network of constraints. *Pattern Recognition Letters*, 22 (2) p.231-248.

[7] Horna, S. et al., (2007). Building 3D indoor scenes topology from 2D architectural plans. *Computer Graphics Theory and Applications (GRAPP)*, pp. 37-44. Available at: http://citeseerx.ist-.psu.edu/viewdoc/download?doi=10.1.1.100.6752 &rep=rep1& type=pdf.

[8] AutoDesk, *AutoCad Products.*[online] Available at: http://usa.autodesk.com/adsk/servlet/pc/index?siteID=123112&id=15408302 [Accessed 15 Dec 2010]

[9] Pilkinton, B. (2010, 12 1). Article Squeeze .*The Importance Of 3D Architectural Visualisation*. [online] Available at: http://www.articlesqueeze.com/business-articles/the-importance-of-3d-architectural-visualisation/ [Accessed 15 Dec 2010]

[10] *Google Sketchup*. [online] Available at: http://sketchup.google.com/intl/en/index.html[Accessed27 Dec, 2010]

[11] Autodesk. *3DS Max Products.*[online] Available at: http://usa.autodesk.com/adsk/servlet/pc/index?siteID=123112&id=15683479 [ Accessed 27 Dec 2010]

[12] Autodesk .*Autodesk Maya*. [online] Available at: http://usa.autodesk.com/adsk/servlet/pc/index?id=13577897&siteID=123112 [ Accessed 27 Dec 2010]

[13] Blender .*Blender*. [online] Available at: http://www.blender.org/ [ Accessed 27 Dec 2010]

[14] Irrlitch*Irrlitch* . [online] Available at:http://irrlicht.sourceforge.net/ [ Accessed 10 Jan 2011]

[15] Gibson S, Howard T. (2000). Interactive reconstruction of virtual environments from photographs, with application to scene-of-crime analysis. In *Proceedings of ACM Symposium in Virtual Reality Software and Technology 2000*, October 2000. Seoul, Korea

[16] Fisher R et al. (2004). *Explore with Java -Adaptive Thresholding Chapter.*[online] Available at:http://homepages.inf.ed.ac.uk/rbf/HIPR2/adpthrsh.htm[Accessed 24 Dec 2008]

[17] Green B (2002). *Canny edge Detection Tutorial*[online] Available at: http://www.pages.drexel.edu/~weg22/can_tut.html [Accessed 14 March 2009]

[18] Raphael C.. Gonzalez, Richard E. Woods(2002). Digital image Processing, 2nd Edition, Prentice Hall International.

[19] Fisher R et al. *Intensity Histogram*[online] Available at: http://homepages.inf.ed.ac.uk/rbf-/HIPR2/histgram.htm[Accessed 25 Dec 08]

[20] Gary Bradsky , Adrian Kaehler. (2008). Learning OpenCV. O'Reilly.

[21] G.Chaudri.. *Bhattacharyya distance*. Available at: http://eom.springer.de/B/b110490.htm [Accessed 15 March 09]

[22] Longin Jan Latecki. *Template matching*, Available at: http://www.cis.temple.edu/~latecki/Courses/CIS601-03/Lectures/TemplateMatching03.ppt[Accessed 14 March 2009]

[23] Binary Image processing.[Online].Available at http://homepages.inf.ed.ac.uk/rbf/CVonline/-LOCAL_COPIES/MARBLE/medium/binary/ [Accessed 18 March 2009]

[24] OpenCV, Open Computer Vision Library. [Online] Available at: http://sourceforge.net/projects/opencvlibrary/ [Accessed 18 June 2011]

[25] Xuetao Yin, Wonka, P., Razdan, A. Generating 3D Building Models from Architectural Drawings: A Survey, IEEE Computer Graphics and Applications - CGA , vol. 29, no. 1, pp. 20-30, 2009. DOI: 10.1109/MCG.2009.9

[26] Su, F., Song, J. & Cai, S., 2001. Dimension recognition and geometry reconstruction in vectorization of engineering drawings. *Proceedings of the 2001 IEEE Computer Society Conference on Computer Vision and Pattern Recognition CVPR 2001*, p.I-710-I-716

[27] Jung, I., Mckinley, L. & Weg, Z., 2008. Three-dimensional building reconstruction : a process for the creation of 3D buildings from airborne LiDAR and 2D building footprints for use in urban planning and environmental scenario modelling. Available at: http://www.virtualcitysystems.de/uploads/-media/BuildingReconstruction_Paper_RealCorp_2008.pdf

[28] Horna, S. et al., 2009. Consistency constraints and 3D building reconstruction.*Computer-Aided Design*, 41(1), p.13-27

## Authors

**Mr. Raj Kishen Moloo** is a lecturer at the University of Mauritius. He has a Masters degree in Advanced Computer Science at the University of Manchester (2005) and BEng (Hons) Computer Science and Engineering (2003) from the University of Mauritius. He is also an ACCA affiliate. His main research interests are Computer Graphics, Games, Virtual Reality and Mobile Application Development.

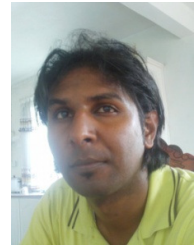

**Mr. Muhammad Ajmal Sheik Dawood** is enrolled for an MSc Financial Software Engineering at the University of Essex. He has a BSc (Hons) Computer Science and Engineering from the University of Mauritius.

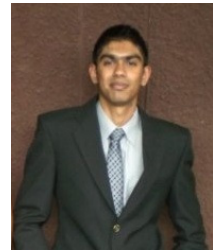

**Mr. Abu Salmaan Auleear** is working as a Senior Web-Designer at Chesteroc Ltd (Mundocom Mauritius). He has a BSc (Hons) Computer Science and Engineering from the University of Mauritius.

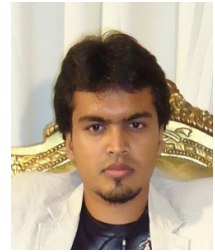